\documentclass[entropy,article,accept,moreauthors,pdftex,10pt,a4paper]{mdpi}
\include{unicode}
\firstpage{1} 
\articlenumber{164}
\doinum{10.3390/e18050164}
\pubvolume{18}
\pubyear{2016}
\copyrightyear{2016}
\externaleditor{Academic Editor: Andreas Holzinger }
\history{Received: 30 January 2016; Accepted: 22 April 2016; Published: 27 April 2016}

\usepackage[caption=false]{subfig}
\usepackage{graphicx}
\graphicspath{{images/}}
\DeclareGraphicsExtensions{.pdf,.eps,.png}
\usepackage{booktabs}
\usepackage{multirow}
\usepackage{amssymb}
\usepackage{tipa}
\usepackage{soul}
 \theoremstyle{mdpi}
 \newcounter{thm}
 \newcounter{ex}
 \newcounter{re}
 \theoremstyle{mdpidefinition}

\Title{Finding Influential Users in Social Media Using Association Rule Learning}

\Author{Fredrik Erlandsson \textsuperscript{1,}*, Piotr Br\'{o}dka \textsuperscript{2}, Anton Borg \textsuperscript{1} and Henric Johnson \textsuperscript{1}}

\address{%
\textsuperscript{1} \quad Blekinge Institute of Technology, Karlskrona 371\,79, Sweden; anton.borg@bth.se {(A.B.); henric.johnson@bth.se (H.J.)} \\
\textsuperscript{2} \quad Wroc\textbarl aw University of Technology, 50-370 Wroc\textbarl aw, Poland; piotr.brodka@pwr.edu.pl}

\corres{Correspondence: fredrik.erlandsson@bth.se; Tel.: +46-455-385-000}

\abstract{Influential users play an important role in online social networks since users tend to have an impact on one other. Therefore, the proposed work analyzes users and their behavior in order to identify influential users and predict user participation. Normally, the success of a social media site is dependent on the activity level of the participating users. For both online social networking sites and individual users, it is of interest to find out if a topic will be interesting or not. In this article, we propose association learning to detect relationships between users. In order to verify the findings, several experiments were executed based on social network analysis, in which the most influential users identified from association rule learning were compared to the results from Degree Centrality and Page Rank Centrality. The results clearly indicate that it is possible to identify the most influential users using association rule learning. In addition, the results also indicate a lower execution time compared to state-of-the-art methods.}

\keyword{social media; data mining; association rule learning; prediction; social network analysis}

\begin{document}

\section{Introduction} 
\label{sec:introduction}
Online social networks are playing an important role in our society and have created a platform for people to communicate and express their thoughts. With the use of online social media, we have created a way to mimic real human communication in an online environment. Facebook alone attracts $1.3$ billion users with 640 million minutes spent each month on the site. Consequently, discovering trending topics or influential users is of interest for many researchers interested in areas such as marketing~\cite{cha2010measuring}. Several studies have tried to identify user influence; however, most have used Page Rank Centrality~\cite{DBLP:journals/corr/Riquelme15,Musial:2009:UPM:1731011.1731017} or Degree Centrality~\cite{Musial:2009:UPM:1731011.1731017,brodka2013key} based approaches to identify influential users. This paper builds on the initial discoveries on association rule learning in social networking sites:~\cite{Erlandsson:2016kq}.

In this article, we argue that users on Facebook groups are following each other and that it is possible to detect influential users and predict user participation. For example, if users A, B, C and D share common interests, there is a chance that if A, B, and C already have commented on a topic, D will also comment on it. Therefore, this paper relates to how users perform actions (e.g., comments or likes) on posts in Facebook pages. In addition, we use association rule learning to discover relationships between users in our dataset~\cite{flach2012machine}. Given a list of posts from a specific domain, we extract users' actions, such as comments and likes. Using association rule learning on the data, we argue that it is possible to predict if a particular user will or will not participate on a post discussion based on the other users' activity.

This article has three major contributions: firstly, possibilities to identify influential users using association rule learning are presented; secondly, we present time performance of well-known methods for ranking users in social media together with our approach using association rule learning; and finally, we show how association rule learning can be used to predict user participation.

For evaluation, several experiments are conducted, which include building association rules that can be used to predict if a specific user will be active in a particular post. The prediction is done based on the activeness of users within current posts. In addition, an extended social network analysis is conducted to verify the findings of influential users.

The paper is organized as follows: in Section~\ref{sec:related_work}, related work is discussed; in Section~\ref{sec:association}, association rule learning and the evaluation metrics are discussed; in Section~\ref{sec:data_model}, the dataset is presented; and finally, the results are presented in Section~\ref{sec:resuts} and discussed in Section~\ref{sec:discussion}.

\section{Related Work} 
\label{sec:related_work}
Online social networks and social media analysis are popular research areas in contemporary network science. The main focus in social network research is on link prediction~\cite{Liben-Nowell:2007:LPS:1241540.1241551} and social connection prediction~\cite{utz2015making}. Different teams around the world also work on: (i) personality prediction for micro blog users~\cite{xia}, (ii) churn prediction and its influence on the network~\cite{au2003novel,ruta2009network}, (iii) community evolution prediction~\cite{saganowski2015predicting,DeMeo:2015cf}, (iv) using social media to predict real-world outcomes~\cite{Asur:2010:PFS:1913793.1914092}, (v) predicting friendship intensity~\cite{Ahmad2010,nia2013leveraging}, (vi) affiliation recommendations\cite{Spertus:2005ec,Vasuki:2011ku}, and (vii) sentiment analysis and opinion mining~\cite{Petz:2015fj} .

Other popular areas of research focus on popularity prediction in social media based on comment mining~\cite{5368318}, predicting information cascade on social media~\cite{7062665}, and predicting patterns of diffusion processes in social network~\cite{Jankowski2012}. An important factor is often the user's role in the different processes. As such, identifying influential users are of interest to understand and/or affect the spread of information, e.g., viral marketing. The ability to identify influential users might also affect the research into other areas of related work (e.g., ii or iii).

Research into detecting influential users on Twitter indicates that, while a large amount of followers seem to be present among influential users, predictions of which particular user will be influential is unreliable~\cite{Bakshy:2011go}. Depending on the social network, how to define influence differs, e.g., influence on Twitter might be defined by retweets or mentions, while, on Digg, votes generated are used to measure influence~\cite{cha2010measuring,Ghosh:2010vz,Shin:2008cz}. While some initial research has been done using clustering algorithms to identify top users, based on influence features, e.g., likes and replies, evaluation is lacking~\cite{Lin:2015kh}. Similarly, linear regression has been used to identify influential (categorical) users based on influence features~\cite{Shin:2008cz}.

While some research on identifying influential users use learning based approaches, another popular approach to identifying influential users is the Page Rank algorithm or adaptions of the Page Rank algorithm~\cite{Weng:2010fd, Tang:2010dh, Hotho:2006bs}.

\citet{Nancy2013} explore the association rule between a course and gender in the Facebook 100 university dataset. This was performed to discover the influence of gender in studying a specific course. \citet{Xiaoqing} introduce the scheme for association rule learning of personal hobbies in social networks, while~\citet{Schmitz2006} tackle the problem of mining association rules in folksonomies and try to find out how association rule learning can be applied to analyze and structure folksonomies.

Initial research used association rule learning to identify influential users and predict user participation in online social networks~\cite{Erlandsson:2016kq}. Association rule learning has been previously used in social network and social media analysis.

While online social network analysis is popular, there is, according to our review, a lack of research on using association rules for predicting user participation in online social media discussions.

\section{Association Rule Learning} 
\label{sec:association}
Association rule learning is a machine learning technique that aims to find out how one item affects another by analyzing how frequently certain items appear together in a specific dataset. This is done by using two criteria, namely, \emph{support} and \emph{confidence}. Support indicates the frequency of such items, while confidence indicates how many times those rules in the whole dataset are correct. An example of an association rule is the following: ``Ninety-percent of transactions that purchase bread and butter also purchase milk''~\cite{Agrawal:1993:MAR:170036.170072}.

As stated in Section~\ref{sec:introduction}, we are trying to assess user participation in a post based on previous interactions with other users on common posts within one page. We assume that if user A participates in most of the posts where user B is participating as well, there is a high chance of A participating in a new post where B is already active, either because participation of B influences A to participate and/or they both have similar interests. The method of matching items in different transactions is called association rule learning. We apply association rule learning to the domain of social media where we model the data as follows. Items correspond to users on Facebook and transactions correspond to posts. A user is considered to be active and part of the transaction as an item if the user comments on a post.

From the selected dataset described in Section~\ref{sub:data_selection}, we firstly count the frequency of all posts where A and B are active, respectively. Secondly, we count all posts where $A \cup B$ both participate. This gives us two measures, length (the number of participating users in the set) and frequency (the sum of all posts where the users are participating). These two steps can be summarized as building frequent item-sets ($\{\mathcal{I}\}$). Finally, all possible rules from the computed $\{\mathcal{I}\}$s are generated. In this step, we also compute the evaluation metrics described below.

\subsection{Evaluation Metrics}
\label{sub:evaluation_metrics}
Several metrics exist that will help understand the learned association rules. The first measure, \textit{Support}, shows how big of a portion of $\{\mathcal{D}\}$ the item-set covers. It is calculated by dividing the frequency of a given item-set, $\{\mathcal{I}\}$, with the total number of transactions (posts) in our dataset, $\{\mathcal{D}\}$, or the number occurrences of $\{ A,\,B\}$ divided by the number of items in $\{\mathcal{D}\}$. As shown in Equation~(\ref{eq:support}):

\begin{equation}\label{eq:support}
	support\big(\{A,B\}\big) = \frac{\{A,B\}}{|\mathcal{D}|}.
\end{equation}

The second measure, \textit{Confidence}, indicates the proportions of transactions that contain $\{A,\,B\}$ that also will contain $C$  in the set of transactions in $\{\mathcal{D}\}$, given the following rule $\{A,\,B\} \Rightarrow C$. \textit{Confidence} is calculated as shown in~Equation~(\ref{eq:confidence}).
Say that $\{A,\,B,\,C\}$ participates in four common posts and $\{A,\,B\}$ participates in eight posts in total. This leads to $4 / 8 = 0.5$, or the \textit{confidence} that $C$ will participate on a post where $A$ and $B$ already are active is $50\,\%$:

\begin{equation}\label{eq:confidence}
	con\hspace{-0.1em}f\hspace{-0.1em}idence\big(\{A,\,B\} \Rightarrow C\big) = \frac{support(\{A,\,B,\,C\})}{support(A,\,B)}.
\end{equation}

The third measure, \textit{lift}, shows the ratio of interdependence of the observed values. As we see from~Equation~(\ref{eq:lift}), if lift is 1, it implies that the rule and the items are independent from each other. However, if lift is $>\,1$, the lift indicates the dependency of our item-sets:

\begin{equation}\label{eq:lift}
	li\hspace{-0.1em}f\hspace{-0.1em}t\big(\{A,\,B\} \Rightarrow C\big) =  \frac{support(\{A,\,B,\,C\})}{support(\{A,\,B\}) \times support(\{C\})}.
\end{equation}

Finally, \textit{conviction} is the ratio of the expected \textit{support} that $\{A,B\}$ occurs without $C$ as shown in~Equation~(\ref{eq:conviction}). Notably,  \textit{conviction} is infinite (due to division with zero) when the \textit{confidence} is 1:

\begin{equation}\label{eq:conviction}
	conviction\big(\{A,\,B\} \Rightarrow C \big) =
  \frac{1- support(\{A,\,B\})}{1-con\hspace{-0.1em}f\hspace{-0.1em}idence\big( \{A,\,B\} \Rightarrow C\big)}.
\end{equation}

The described measures enable understanding of the learned rules in $\{\mathcal{D}\}$, where higher numbers of all four measures indicate that the learned rule has relevance for prediction.

\subsection{Usage of the Eclat Algorithm}
\label{sec:methodology}
To build association rules from our dataset, we evaluated several implementations. \citet{Agrawal:1994:FAM:645920.672836} presented the Aprori algorithm, which was proven to be an efficient method for association rule learning. However, this algorithm is proven to have efficiency issues in large datasets~\cite{Goethals02surveyon}, and the identified implementation for Python is very slow (considering that in our dataset it was not possible to get a result within a reasonable time). Hence, other algorithms were tested, in particular, the Eclat algorithm~\cite{10.1109/69.846291}. The Eclat algorithm quickly discards items with low frequency by considering a minimum number of associations as input parameters. We have found that a reasonable trade-off between resolution and speed is four, in our dataset, where a lower frequency of items is ignored. The use of four as a lower bound was identified empirically by starting at the number of comments divided by the number of users and then calculating the item-sets with decreasing threshold until the execution speed reached 10 s. At 10 s, all available RAM memory in our experiment environment was exhausted, and we stopped the execution. For one of the investigated pages, we saw that with a threshold of five, we can generate 4230 item-sets in 350 ms, and with a threshold of four, we can generate 9117 item-sets in 600 ms. A threshold of three fills up available resources and {never completes} the calculations.

\section{Data Model}
\label{sec:data_model}
The data used in this study have been obtained from the crawler described by \mbox{\citet{erlandsson2015crawling}}. This crawler gathers complete posts from Facebook. In this context, the term complete, stands for posts that contain all likes and comments created up to the crawling time as well as the data about the users who have created them.
Our current dataset, captured from public pages and groups on Facebook, consists of over 56\,million posts, 560\,million comments and $7.3$\,billion likes made by 820\,million Facebook users. The crawled data was parsed and made available from an SQL database, structured as described in~\cite{nia2012sin}, making all fields needed for our task available. In this study, we assume that the investigated posts will not get any new comments. We simplify the dynamics of social media by saying that the posts we are investigating were ``dead'' when the data was collected, in which the term of dead posts refers to posts that no longer attract attention, new comments, or~likes.

This study is limited to only active users. Thus, we exclude posts with less than 20 comments and users who had less than five comments, as they are considered to be occasional visitors and not real page participants.

\subsection*{Data Selection}
\label{sub:data_selection}
We have sampled 195 pages from our dataset, varying in terms of the number of users, posts, comments and user activity to make the sample of Facebook data as broad and as diverse as possible.
Despite the fact that we have calculated the rules using a server with 144\,GB of RAM memory and a 24 core processor, we could not calculate the rules for the biggest pages (44 of them), thus we had to remove them from our dataset. An example of such a page is \textit{Fox News} with 837,176 users 4485 posts, 6,967,304 comments, and a lifetime of 2034 days (almost six years).
An additional 43 pages had to be removed because they were too small, \emph{i.e}., having less than 10 posts with more than 20 comments and/or less than 10 users with more than five comments.
After the preprocessing, we still had 108 pages ranging from 152 to 675,200 active users, from 18 to 161,264 posts, and from 577 to 1,340,730 comments. Table~\ref{tab:dataset} presents the descriptive statistics of this dataset.

For the initial results, the page \cite{OccupyTogether} has been selected. This page was selected based on the following properties: it is active, it has a high number of users, and it is political with a biased user group (most of the users have positive perceptions of the Occupy movement). It was also selected as it is a page in the median range of the complete dataset with respect to the number of active users, 2443, and active posts, 610.
\begin{table}[H]
\centering
\setlength{\tabcolsep}{4pt}
\caption{Filtered descriptive statistics of the dataset of 108 pages.}\label{tab:dataset}
\begin{tabular}{lrrrrrrr}
\toprule

\textbf{Type}       &   \textbf{Mean} &    \textbf{Std.} &  \textbf{Min} &  \boldmath$Q1$ & \textbf{Median} &     \boldmath$Q3$ &     \textbf{Max} \\

\hline

Users    &  69,678 & 130,564 &  152 & 4282 & 17,995 &  62,194 &  675,200 \\
Posts    &   7431 &  19,329 &   18 &  784 &  2157 &   5758 &  161,264 \\
Comments & 147,721 & 264,711 &  577 & 7886 & 33,437 & 133,421 & 1,340,730 \\

\bottomrule
\end{tabular}
\end{table}
\section{Experiments and Results}
\label{sec:resuts}
To verify the findings, several experiments were executed. These experiments were firstly performed on the page Occupy\-Together, and were extended to the whole dataset described in Section~\ref{sub:data_selection} for verification of the results. First, a comprehensive experiment of association rule learning was conducted.  Secondly, the learned rules were evaluated with respect to prediction accuracy of user participation using a training test split ($80/20$). Finally, social network analyses for each page were performed to verify and evaluate ranked users identified as influential by the first experiment.

\subsection{Item-Sets and Rules}
\begin{figure}[H]
  \centering
  \includegraphics[width=0.45\textwidth]{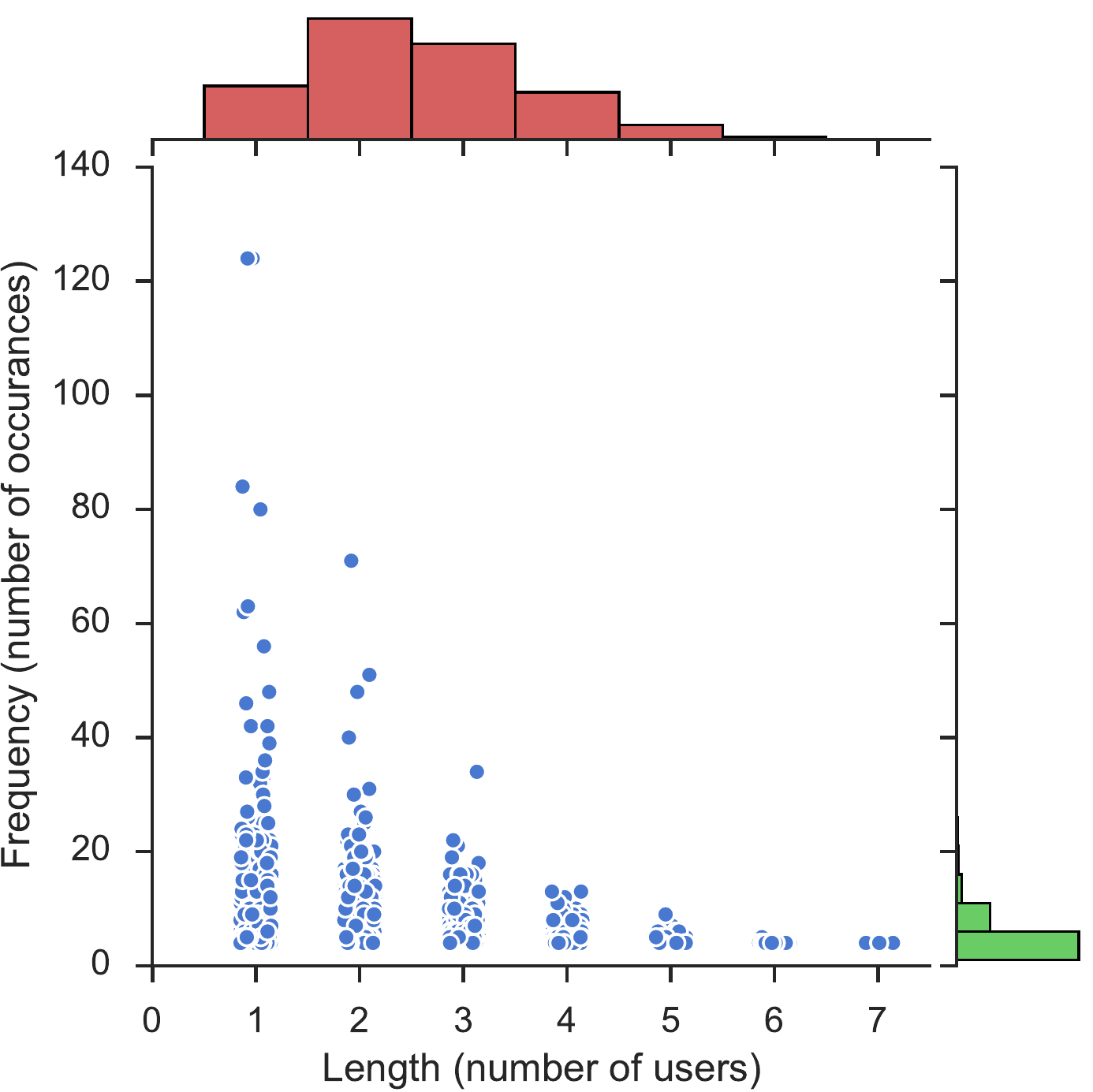}
  \caption{Combined plot of number of occurrence of each item-set (Frequency) with respect to  number of users in the rule (Length). The upper and right axis illustrates histograms of the respective~distributions.}\label{fig:userParticipation}
\end{figure}

Using the methods described in Section~\ref{sec:methodology}, an experiment was performed to create frequent item-sets and build association rules for these sets.
The resulting frequent item-sets are depicted in Figure~\ref{fig:userParticipation} for the page {Occupy\-Together}. This figure illustrates frequency, or the number of occurrences for each item-set, with respect to the length of elements, or the number of collaborating users. The main scatter-plot illustrates how the \emph{frequency} decreases when the number of users (\emph{length}) increases, a natural feature of frequent item-sets.
Figure~\ref{fig:userParticipation} also depicts the distribution as histograms. The top histogram, in green, shows the distribution of frequency and, the histogram on the right hand side, in red, shows the distribution of the length of the learned item-sets. The histogram to the right (in green) illustrates a significant density of user collaboration that occurs at a low frequency, between 1 and 10. This is natural as the frequency of user participation decreases for most of the users. Noticeable on the length distribution (in red) is the fact that the density is higher for two and three participating users than for just one. This is because there exist more combinations of users than the number of single users.

Association rules supporting the hypothesis of user participation based on other users' activities were computed from the calculated frequency item-sets. This resulted in $55,166$ rules for the page~\cite{OccupyTogether}. Table~\ref{tab:stat} shows descriptive statistics for all the computed rules. It can be noted that although the confidence median and mean is low, the high level of lift indicates a high dependency of the learned rules, \emph{i.e}., the computed rules show that our hypothesis is valid and users tend to follow each other. Since our dataset is big, with many users and many posts, a low support mean and median is expected. Moreover, it is noticeable that users are not active in all posts but more on a subset of them.

\begin{table}[H]
\centering
\setlength{\tabcolsep}{4pt}
\caption{Descriptive statistics of $55,166$ computed rules.}\label{tab:stat}
\begin{tabular}{lccc}
\toprule

 \textbf{Evaluation Metric} & \textbf{Mean} & \textbf{Median} & \textbf{Std.} \\

\hline
Support & 0.05 & 0.02 & 0.07 \\
Confidence & 0.43 & 0.33 & 0.33 \\
Lift & 18.97 & 9.38 & 24.64 \\
Conviction & 1.83 & 1.32 & 1.18 \\
\bottomrule
\end{tabular}
\end{table}

\begin{figure}[H]
  \centering
  \subfloat[]
  {\includegraphics[width=0.48\textwidth]{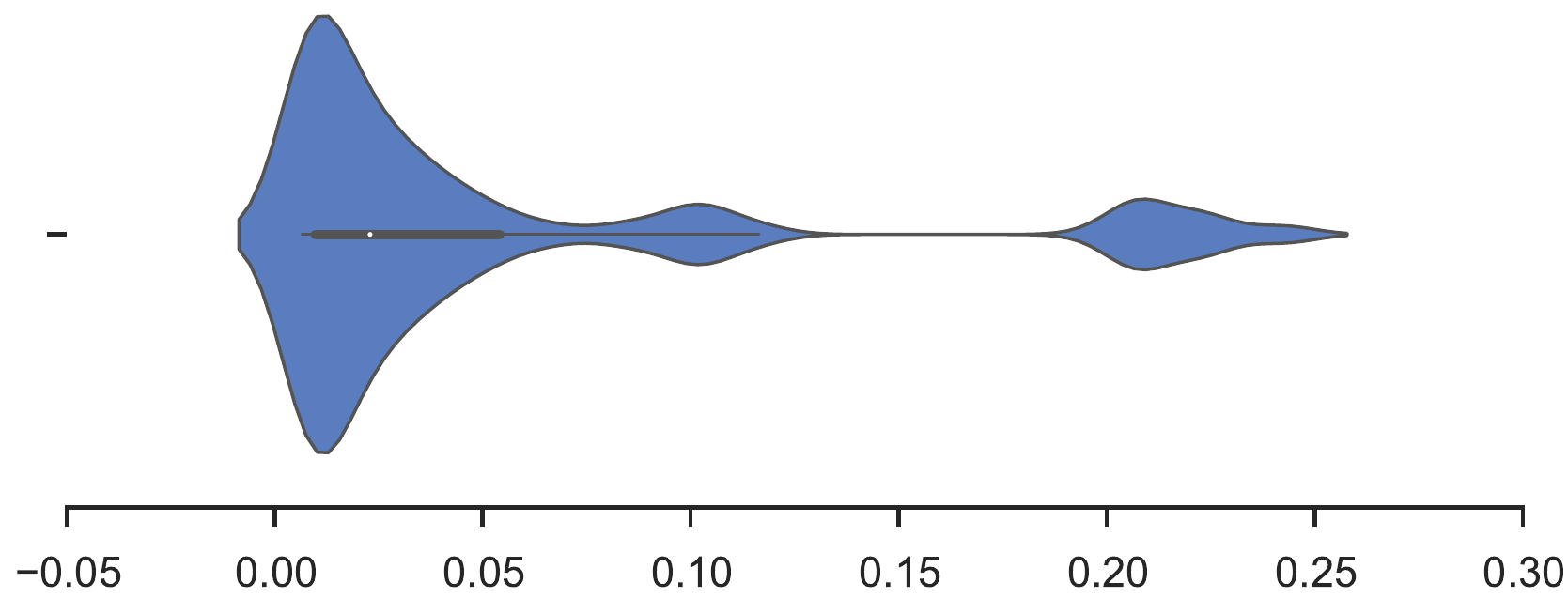}\label{fig:violinplot_support}}
  \quad
  \subfloat[]
  {\includegraphics[width=0.48\textwidth]{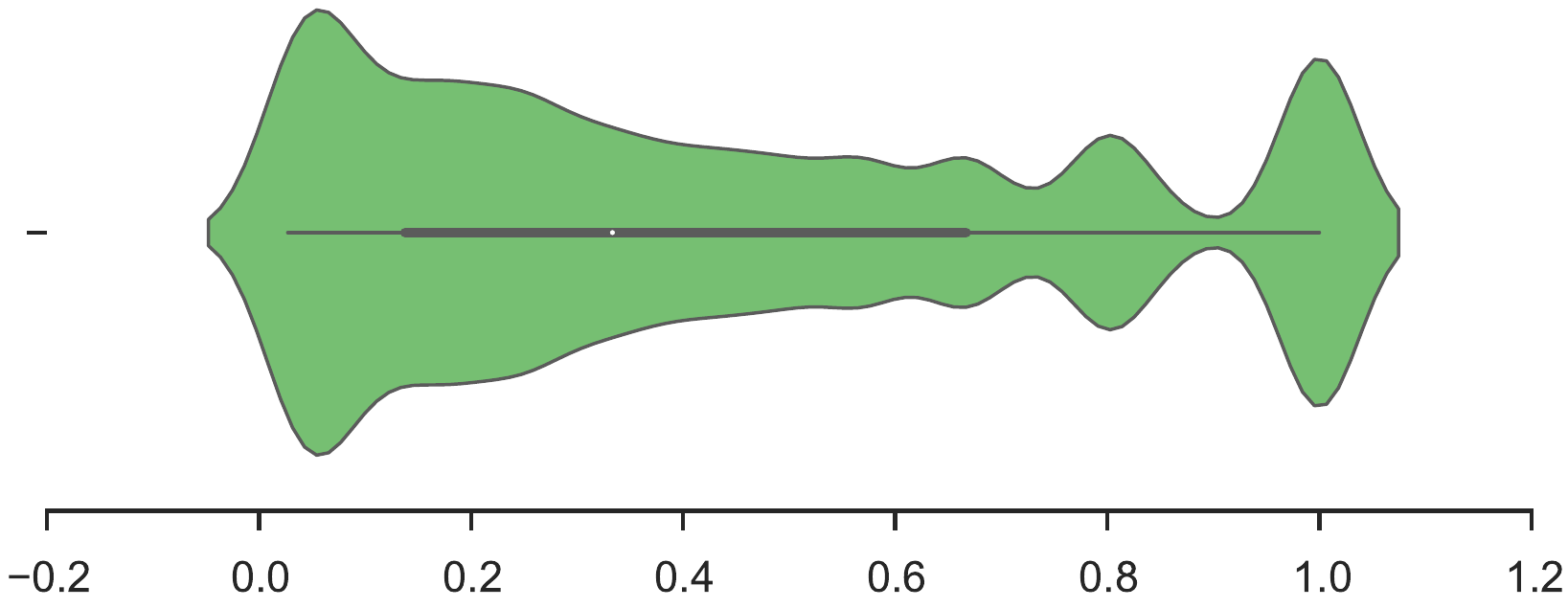}\label{fig:violinplot_conf}}
  \\
  \subfloat[]
  {\includegraphics[width=0.48\textwidth]{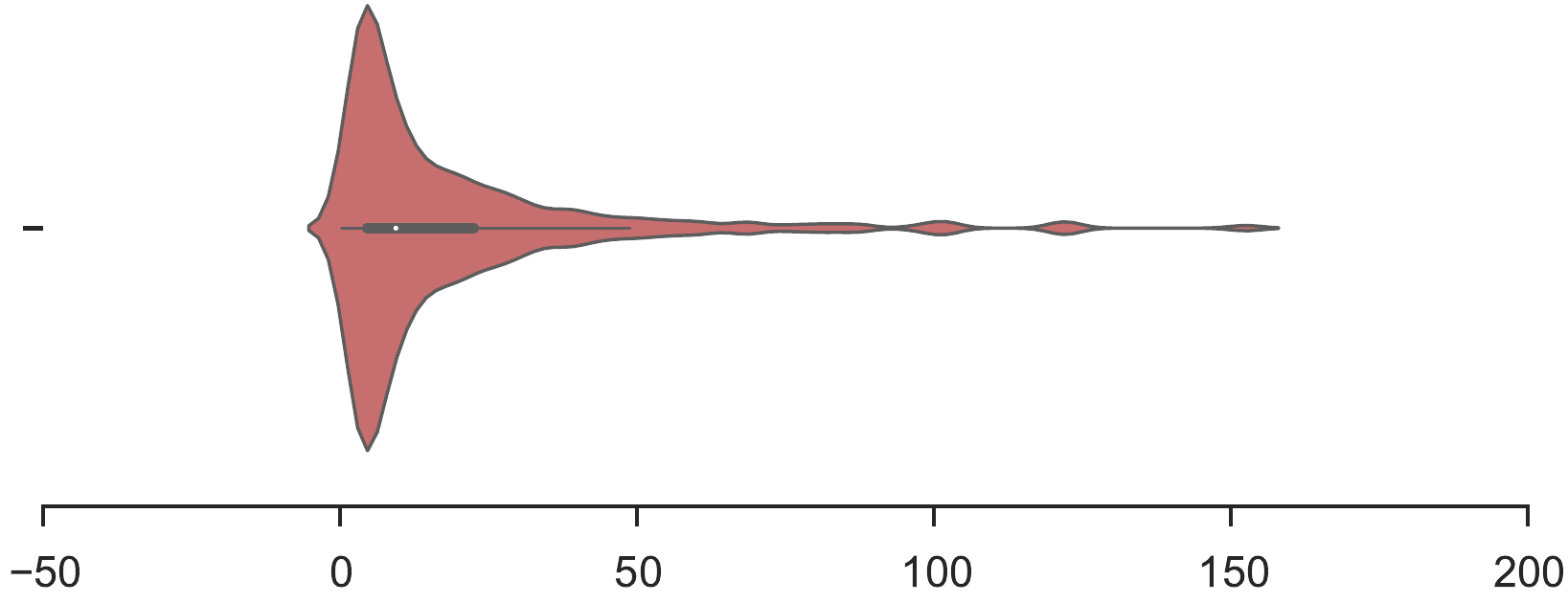}\label{fig:violinplot_lift}}
  \quad
  \subfloat[]
  {\includegraphics[width=0.48\textwidth]{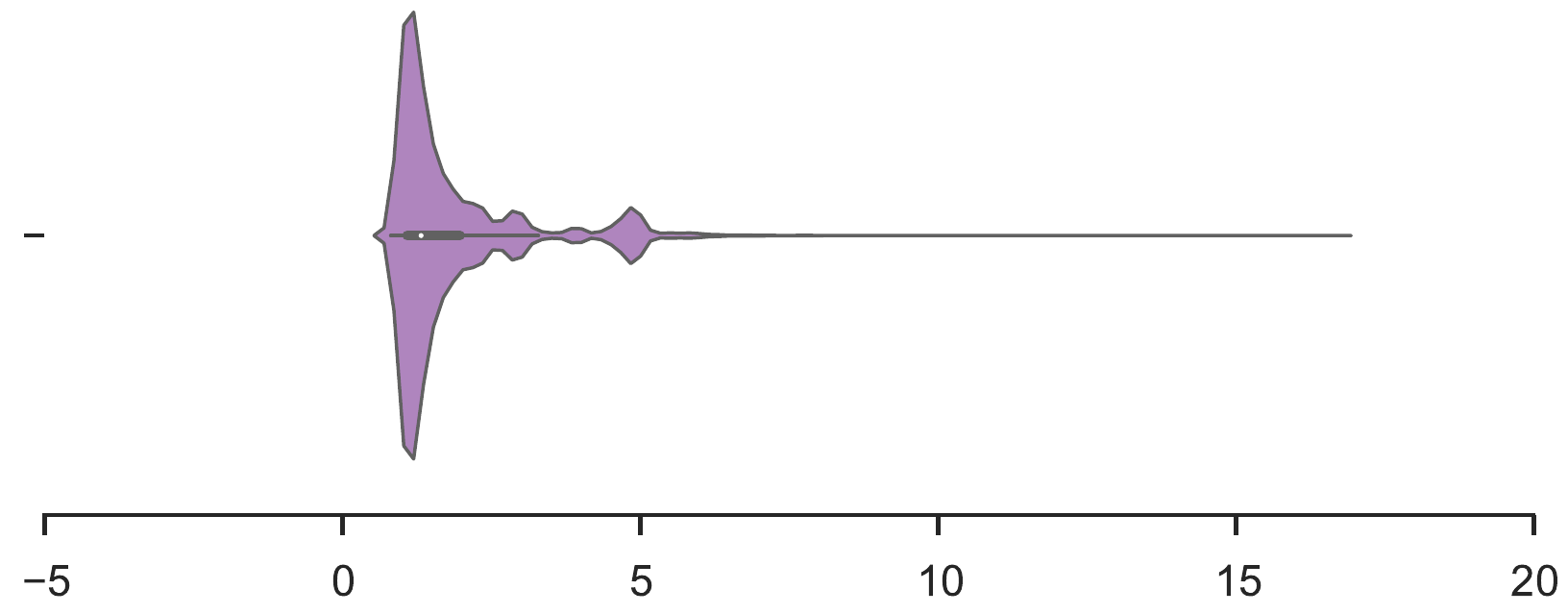}\label{fig:violinplot_conv}}
  \caption{Distribution of values in learned association rules. (\textbf{a}) support distribution; (\textbf{b}) confidence distribution; (\textbf{c}) lift distribution; (\textbf{d}) conviction distribution.}
  \label{fig:violinplots}
\end{figure}

Figure~\ref{fig:violinplots} depicts the distribution, Confidence, Lift, Conviction and Frequency respectively in our learned model. The figures are violin-plots, which illustrate the kernel density (shown as height and depth) in addition to normal box-plots with outer quartiles as thin lines,  inner quartiles as bold lines and the mean as a white dot.

Figure~\ref{fig:violinplot_support} shows a dense distribution of support at $0.025$ and, interestingly, a higher density at $0.20$. The confidence distribution is illustrated in Figure~\ref{fig:violinplot_conf}, in which we obtained a dense distribution around $1.0$, \emph{i.e}., there are a significant number of learned rules with high confidence, thus, the rule is accurate. Figure~\ref{fig:violinplot_lift} shows that the lift measure has a heavy tail distribution. In addition, Figure~\ref{fig:violinplot_conv} illustrates a distribution of conviction to be concentrated between zero and five.

Table~\ref{tab:rules} presents learned rules in three sections. Each section is sorted firstly, by Confidence, Lift and Conviction, respectively, and secondly by the number of supporting users. The rule $\{u_{429},\ u_{578}\}\ \Rightarrow\ \{u_{19}\}$ should be interpreted as user 429 together with user 578 influencing the participation of user 19. Notably, when sorting by confidence and lift, the conviction is infinite (this is due to the confidence of $1.0$) which is shown in how conviction is calculated in~Equation~(\ref{eq:conviction}). All of the rules in Table~\ref{tab:rules} have high confidence and show high dependency (via the lift metric), \emph{i.e}., the top five rules sorted by either Confidence, Lift or Conviction are relevant for predicting user participation.

The rule, $\{u_{580},\ u_{861},\ u_{1352},\ u_{1466}\}\ \Rightarrow\ \{u_{896},\ u_{1291}\}$ presented in Table~\ref{tab:rules} with a confidence of $1.0$ and a lift of $152.5$, strongly indicates that the left-hand-side user set influences the right-hand-side user set, \emph{i.e}., when the left-hand-side user set is active on a post, the right-hand-side user set also will be active. A confidence of $1.0$ means that $100\%$ of the posts where the left-hand-side user set is active, the right-hand-side user set also will be active. A lift value of $152.5$, in this specific rule, shows that the right-hand-side user set is dependent on the left.

\begin{table}[H]
\caption{Top 5 rules sorted by different metrics for the Facebook page OccupyTogether.} \label{tab:rules}
\centering
\begin{tabular}{p{1.3em}lcccc}
\toprule
\noalign{\smallskip}
 & \textbf{Rule} & \textbf{Confidence} &\textbf{ Lift} & \textbf{Conviction} \\
\noalign{\smallskip}
\midrule
\multicolumn{2}{l}{ Confidence }\\
& $\{u_{179},\ u_{538},\ u_{580},\ u_{938},\ u_{992},\ u_{1090}\}\ \Rightarrow\ \{u_{11}\}$ & 1.00 & 10.17 & $\infty$ \\
& $\{u_{11},\ u_{31},\ u_{80},\ u_{179},\ u_{992},\ u_{1093}\}\ \Rightarrow\ \{u_{580}\}$ & 1.00 & 4.80 & $\infty$ \\
& $\{u_{11},\ u_{31},\ u_{179},\ u_{580},\ u_{992},\ u_{1093}\}\ \Rightarrow\ \{u_{80}\}$ & 1.00 & 9.53 & $\infty$ \\
& $\{u_{11},\ u_{179},\ u_{538},\ u_{580},\ u_{938},\ u_{953}\}\ \Rightarrow\ \{u_{429}\}$ & 1.00 & 4.84 & $\infty$ \\
& $\{u_{179},\ u_{1094},\ u_{1096},\ u_{1113},\ u_{1171},\ u_{1352}\}\ \Rightarrow\ \{u_{1378}\}$ & 1.00 & 101.67 & $\infty$ \\
\noalign{\smallskip} \midrule
\multicolumn{2}{l}{ Lift }\\
& $\{u_{580},\ u_{861},\ u_{1352},\ u_{1466}\}\ \Rightarrow\ \{u_{896},\ u_{1291}\}$ & 1.00 & 152.50 & $\infty$ \\
& $\{u_{580},\ u_{861},\ u_{1291},\ u_{1352}\}\ \Rightarrow\ \{u_{896},\ u_{1466}\}$ & 1.00 & 152.50 & $\infty$ \\
& $\{u_{31},\ u_{80},\ u_{179},\ u_{580}\}\ \Rightarrow\ \{u_{11},\ u_{992},\ u_{1093}\}$ & 1.00 & 152.50 & $\infty$ \\
& $\{u_{19},\ u_{64},\ u_{673},\ u_{685}\}\ \Rightarrow\ \{u_{54},\ u_{581}\}$ & 1.00 & 152.50 & $\infty$ \\
& $\{u_{580},\ u_{861},\ u_{1291},\ u_{1466}\}\ \Rightarrow\ \{u_{896},\ u_{1352}\}$ & 1.00 & 152.50 & $\infty$ \\
\noalign{\smallskip} \midrule
\multicolumn{2}{l}{ Conviction }\\
& $\{u_{429},\ u_{578}\}\ \Rightarrow\ \{u_{19}\}$ & 0.95 & 3.93 & 16.66 \\
& $\{u_{920}\}\ \Rightarrow\ \{u_{179}\}$ & 0.95 & 4.27 & 16.32 \\
& $\{u_{929}\}\ \Rightarrow\ \{u_{179}\}$ & 0.95 & 4.26 & 15.54 \\
& $\{u_{580},\ u_{1093}\}\ \Rightarrow\ \{u_{179}\}$ & 0.94 & 4.22 & 13.21 \\
& $\{u_{580},\ u_{938}\}\ \Rightarrow\ \{u_{179}\}$ & 0.94 & 4.22 & 13.21 \\
\bottomrule
\hline
\end{tabular}
\end{table}

Considering rules where at least two separate users affect another user with a confidence of \linebreak$\geqslant$95\%, we can reduce the $55,166$ rules to $4959$ rules, which have a median lift of $4.80$ and a median support of $0.21$. In other words, we have close to $5000$ rules that strongly indicate that users are affected by each other when it comes to participating in online social networks. From learned rules, we can also identify influential users, or the users that exists on the left side of multiple rules as presented in Section~\ref{sec:sna}.

The learned rules of the complete dataset are presented in Table~\ref{tab:rulesAll}, after filtering out rules with \linebreak[4]\emph{Confidence}~$\geqslant$95\%.

\begin{table}[H]
\caption{Descriptive statistics of learned rules with of \emph{Confidence} $\geqslant$95\% from the complete dataset. } \label{tab:rulesAll}
\centering
\begin{tabular}{lrrrrrrr}
\toprule

{\textbf{Evaluation Metric}} &       \textbf{Mean} &         \textbf{Std.} &    \textbf{Min} &      \boldmath$Q1$ &       \textbf{Median} &        \boldmath$Q4$ &          \textbf{Max} \\

\hline

No. of rules              &  33,426.89 &   87,457.39 &  2.00 &  151.00 &  2351.00 &  32,053.50 &   724,510.00 \\
Confidence         &      1.00 &       0.00 &  0.97 &    1.00 &     1.00 &      1.00 &        1.00 \\
Lift               &     38.06 &      42.14 &  1.41 &   10.86 &    25.34 &     47.91 &      217.53 \\
Conviction         &     19.39 &       4.61 &  5.88 &   18.07 &    19.79 &     20.70 &       29.46 \\

\bottomrule

\end{tabular}
\end{table}

\subsection{Verification of Learned Rules} 
\label{subsec:testing}

To test how well association rule learning works for predicting user participation, a split, learn and test pattern have been used. For the page in question, we sort all comments based on creation time and use the first 80\,\% for learning and the last 20\,\% of the posts for testing. The learning part is performed as described in Section~\ref{sec:methodology}, and the testing part is carried out as follows: for each post with comments in the testing set, the active users are considered by finding rules that affect the users with respect to temporal order. Say that user $D$ is commenting on a post (in the testing set), and there exists a rule saying that $A, B\ \&\ C$ affect user $D$, this rule will only be considered to be valid if all of $A, B\ \&\ C$ have made at least one comment each before $D$ makes a comment. Of the 787 intersecting users between the learning and test sets, it is possible to predict 113 (14.36\,\%) users, making use of 5310 (9.63\,\%) of the original 55,166 rules.

To calculate \emph{accuracy} and \emph{precision} of learned rules, we have defined \emph{true/false positive/negatives} as follows: A \emph{true positive} is a rule that predicts user activeness, and the user is active. A \emph{false positive} is when a rule predicts user activeness, but the user is not active. A \emph{true negative} is when no user is active, and there is no rule. A \emph{false negative} is when a user is active, but there is no rule. An example of all four classes are shown in Table~\ref{tab:tp}.

\begin{table}[H]
\centering
\caption{Example of false positives and false negatives. Capital letters indicates users and $P_{1-4}$ corresponds to different posts.}\label{tab:tp}
\begin{tabular}{l c r}
\toprule

 \multicolumn{3}{c}{\textbf{Example rule:} \boldmath$\{A,\ B,\ C\}\Rightarrow\{D\}$} \\

\midrule
$P_1=\{A,\ B,\ C,\ D\}$ & $\longrightarrow$ & true positive\\
$P_2=\{A,\ B,\ C\}$ & $\longrightarrow$ & false positive\\
$P_3=\{F,\ G,\ H\}$ & $\longrightarrow$ & true negative\\
$P_4=\{D,\ E\}$ & $\longrightarrow$ & false negative\\
\bottomrule
\end{tabular}
\end{table}

For the page {Occupy\-Together}, an accuracy of $0.886$, precision of $0.291$, and recall of $0.071$ was calculated, with a testing time of 9175\,s. This result is quite low since all learned rules are being considered. To portray a more realistic view of user influence, the  rules were limited to only consider rules with confidence $\geqslant95\,\%$ and rules affecting a single user. Rules affecting more than one user are already covered by the rules affecting a single user, reducing the number of learned rules from 46,170 to 4469 and the execution time down to 890\,s. Showing an accuracy of $0.927$, precision of $0.794$, and recall of $0.017$. The testing was also performed on the rest of the pages and the results are reported in Table~\ref{tab:testResults}.
The recall is low because there are many \emph{false negatives} (calculated with TP/(TP+FN)). The relatively high accuracy is then achieved with a relatively high number of \emph{true negatives} used in (TN+TP)/(TP+FP+TN+FN).
In general, the unfiltered rules show a lower accuracy, precision, and recall compared to the filtered rules. Furthermore, the complexity of the rule set is reduced by filtering the rules, indicating the beneficial use of rule filtering. The rules set was on average reduced by approximately $93\%$. A less complex rules set could be easier to test and also to understand.

\begin{table}[H]
\centering
\caption{Testing of learned \protect\scalebox{.95}[1.0]{rules based on a $80/20\,\%$ learn and test split. SD stands for standard deviation}.}\label{tab:testResults}
\begin{tabular}{lrrrr}
\toprule

 \textbf{Evaluation Metric} &  {\textbf{OccupyTogether}} & {\textbf{OccupyTogether} \boldmath$^a$} &  {\textbf{All pages (SD)}} & {\textbf{All pages} \boldmath$^a$ (SD)} \\
\hline

No. of rules      &     46,170 &  4469 & 99,237 (248,968) &  7092 (14,965) \\
Accuracy   &     0.886 & 0.927 & 0.858 (0.135) & 0.906 (0.128) \\
Precision  &     0.291 & 0.794 & 0.286 (0.287) & 0.633 (0.343) \\
Recall     &     0.071 & 0.017 & 0.138 (0.193) & 0.165 (0.258) \\

\bottomrule
\multicolumn{5}{c}{\footnotesize $^a$ Reduced set of rules limited by having \emph{Confidence} $\geqslant$95\% and only affected one user.} \\

\end{tabular}
\end{table}

\subsection{Identifying and Verifying Influential Users Using Social Network Analysis} 
\label{sec:sna}

The state-of-the-art method for identifying influential users is social networks analysis (SNA), using the methods Page Rank Centrality~\cite{Musial:2009:UPM:1731011.1731017} or Degree Centrality~\cite{brodka2009performance} for ranking users. It is of interest to see how well influential users identified using association rule learning (ARL) match the state-of-the-art techniques.
Therefore, we have conducted an SNA of our pages as follows: for each page, we have created social networks in such a way that two users are linked together if they commented on the same post: next, for all social networks, Page Rank~\cite{Musial:2009:UPM:1731011.1731017} and Degree~\cite{brodka2009performance} measures have been calculated; and, based on those measures, two ordered (descending) user lists were created, one for each of them.

We have created a similar list for the most influential users from association rule learning. Most influential users are defined as the top-\emph{k} users from the left side of the rules, with a confidence level of greater than 95\,\%, that affect other users to comment on posts. In the most influential users list, users are ranked based on how often they appear on the left side of the rule, e.g., if user \(A\) has appeared three times in all rules and users \(B, C\) and D have appeared one, five and four times, respectively, and the list will look as follows: $[C,\ D,\ A,\ B]$.

Finally, we compared the most influential users identified from association rule learning with top users according to the degree and Page Rank.
Comparison between association rule learning, Degree and Page Rank are considered the top 1\,\%, 5\,\%, 10\,\%, 25\,\%, 50\,\%, 75\,\%, and 100\,\% of the most influential users identified by association rule learning, respectively. The comparison was made as an intersection of two sets created from two lists. For example, if the top four users are $[A,\ B,\ C,\ D]$ for Degree and $[F,\ A,\ C,\ D]$ for association rule learning, the intersection of those two sets will be $[A,\ C,\ D]$ and the size of that set is three, and, in this case, the similarity is 75\,\%.

The example of the SNA analysis for one of the pages \cite{OccupyTogether} is presented in Table~\ref{tab:influential}. The table shows that for the top 209 users on the page Occupy\-Together (the 50\,\% most influential users from association rule learning), there is a similarity of 95\,\% between the users ranked by Page Rank and Degree. When considering users ranked from association rule learning, there is a similarity of 51\,\% compared to Degree and 53\,\% compared to Page Rank.

\begin{table}[h]
\centering
\setlength{\tabcolsep}{6pt}
\caption{Comparison of similarity of influential users for the page Occypy\-Together.}\label{tab:influential}
\begin{tabular}{lcccc}
\toprule

{\textbf{Percent of Top Users}}	& \textbf{Users} &	\textbf{Degree\ $\cap$ ASR}  		&	\textbf{Page Rank $\cap$ ASR}  	&	\textbf{Page Rank $\cap$ Degree}\\

\hline
 1\,\%  &           4 &                              0.75 &                                 0.75 &                        0.75 \\
 5\,\%  &          20 &                              0.45 &                                 0.45 &                        0.95 \\
10\,\%  &          41 &                             0.488 &                                0.512 &                       0.927 \\
25\,\%  &         104 &                             0.462 &                                 0.49 &                       0.923 \\
50\,\%  &         209 &                             0.512 &                                0.526 &                       0.947 \\
75\,\%  &         313 &                             0.502 &                                0.556 &                        0.92 \\
100\,\% &         418 &                             0.517 &                                0.565 &                       0.928 \\
\bottomrule
\end{tabular}
\end{table}

\begin{figure}[H]
  \centering
  \includegraphics[width=0.65\textwidth]{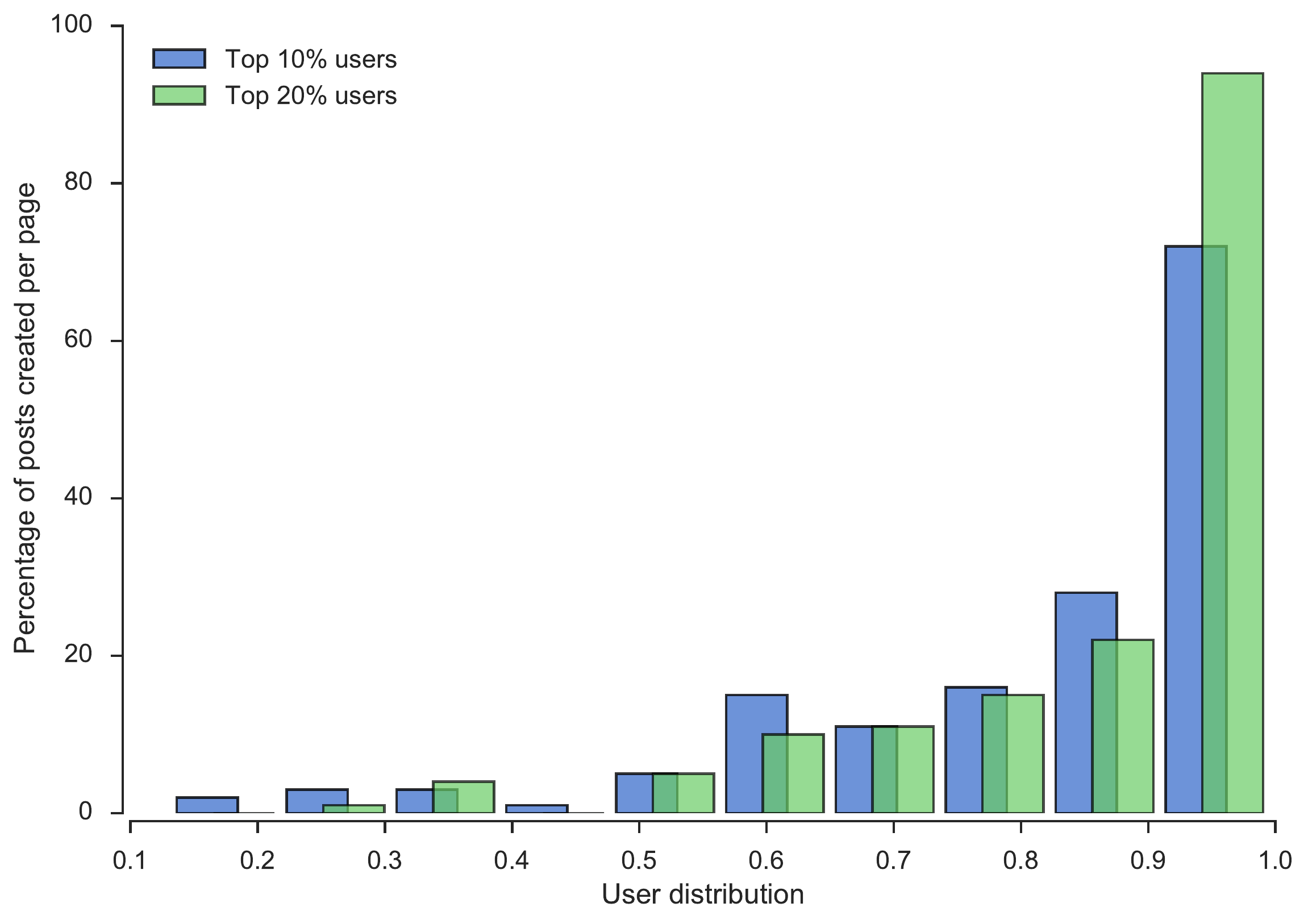}
  \caption{Distribution of posts created by top users over 108 sampled pages.}\label{fig:postHistogram}
\end{figure}

From the SNA analysis, we detected yet another interesting insight into users' behavior in social media pages. We noticed that 10\,\% of users with the highest value of degree measure, created an average of 82.64\,\% posts, and an additional 10\,\% of the most important users add only four more percentage points of posts, \emph{i.e}., 20\,\% of users with the highest value of the degree measure, create 86.84\,\% posts on average.
In Figure~\ref{fig:postHistogram}, the distribution of that phenomena is depicted for all pages.

As described above, the three different approaches were used to detect the most influential users. The intersection between the different user lists were then calculated to evaluate how much each method differs from the others. To detect whether any statistical significant difference exists, Friedman's test was used with the Nemenyi \textit{post hoc} test. Friedman's test is a non-parametric statistical test that ranks the methods over datasets~\cite{Sheskin:2007tx}. When a normal distribution cannot be assumed and several datasets are used, Friedman's test has been suggested as preferable when comparing algorithms~\cite{Demsar:2006un}. The Nemenyi \textit{post hoc} test evaluates between which intersections a significant difference exists. The means and standard deviation for the intersections of several posts are presented in Table~\ref{tab:means}. A low standard deviation indicates that the expected value, \emph{i.e}., the intersection between two sets, is close to the mean. However, there might still exist results which are not close to the mean, e.g., as seen in Table~\ref{tab:influential}.

The average shows that, regardless of the size of the intersection, \emph{Page Rank $\cap$ Degree} has more users in common than the other intersections, while Page Rank and Degree, considered state-of-the-art, have a high amount of users in common (see \emph{Page Rank $\cap$ Degree} in Table~\ref{tab:means}), the rule based learner has fewer users in common with both the Page Rank (\emph{Page Rank $\cap$ ARL}) method and the Degree method (\emph{Page Rank $\cap$ Degree}).

\begin{table}[H]
\centering
\caption{Average intersection measurement and average rank using Friedman's test.}
\label{tab:means}
\begin{tabular}{l c c c}
\toprule

\textbf{Percent of Top Users}	&	\textbf{Degree\ $\cap$ ASR (SD)} 		&	\textbf{Page Rank\ $\cap$ ASR (SD)} 	&	\textbf{Page Rank $\cap$ Degree\ (SD)} \\
\hline

1\,\%	 	&	0.092 (0.173) 	&	0.131 (0.227)	&	0.822 (0.238) \\
5\,\%	&	0.081 (0.145)	&	0.095 (0.158)	&	0.805 (0.251) \\
10\,\%		&	0.115 (0.158)	&	0.133 (0.173)	&	0.830 (0.219) \\
25\,\%		&	0.181 (0.188)	&	0.194 (0.198)	&	0.836 (0.167) \\
50\,\%		&	0.231 (0.212)	&	0.257 (0.228)	&	0.848 (0.129) \\
75\,\%		&	0.266 (0.243)	&	0.286 (0.249)	&	0.868 (0.119) \\
100\,\%		&	0.286 (0.261)	&	0.304 (0.264)	&	0.886 (0.114) \\
\midrule
Average Rank	&	3	&	2	&	1 \\
\bottomrule
\end{tabular}
\end{table}

Friedman's test shows that there are some significant differences between the intersects, \linebreak $\chi^{2}=9.210$, $df=2$, $p=0.01$. The Nemenyi test result (see Table~\ref{tab:nem}) demonstrates that the \emph{Page Rank $\cap$ Degree} set performs significantly better than the \emph{Degree $\cap$ ARL} set at a confidence level of both $0.95$ and $0.99$.

\begin{figure}[H]
  \centering
  \includegraphics[width=0.65\textwidth]{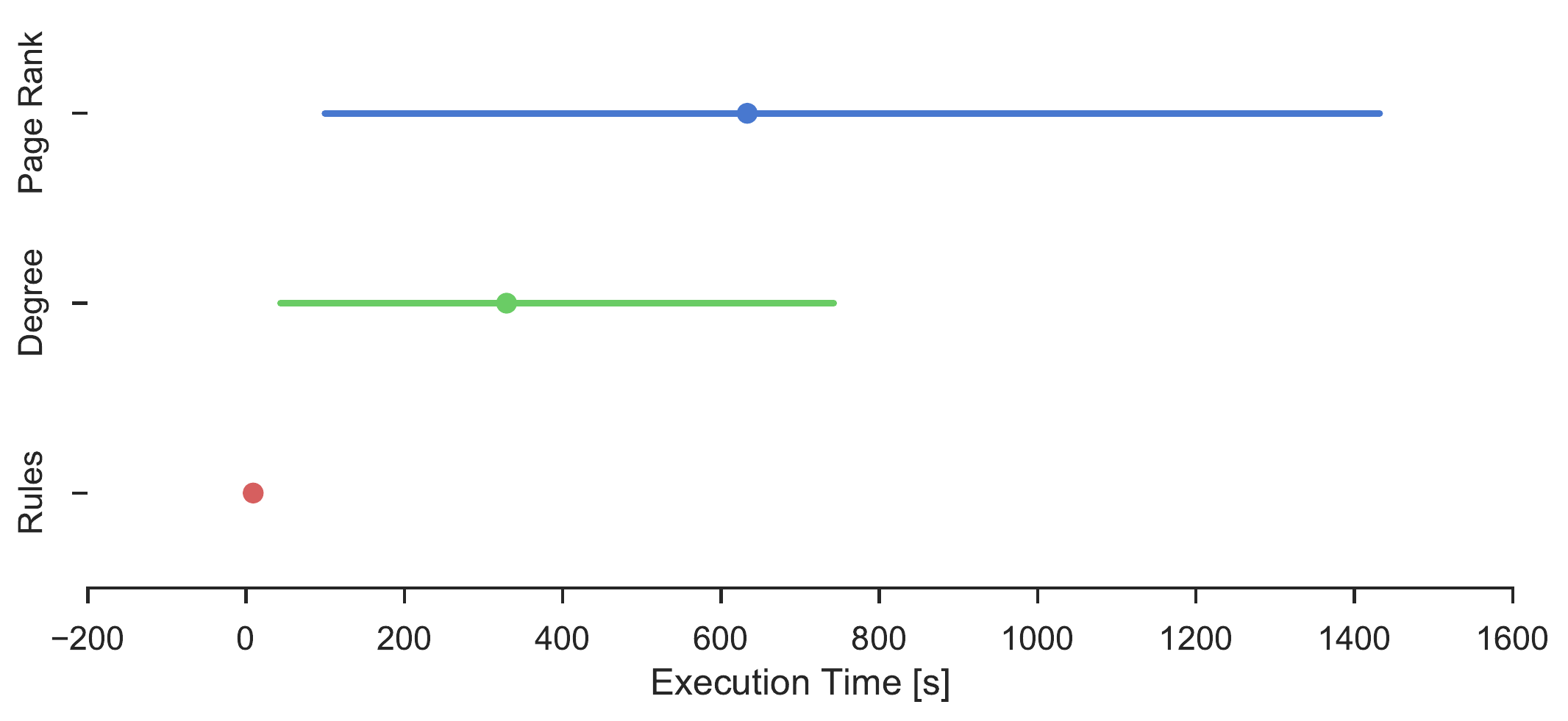}
  \caption{Execution time for different social network analysis methods.}\label{fig:sna_time}
\end{figure}

\begin{table}[H]
\centering
\caption{Paired rank comparison of intersections using the Nemenyi \textit{post hoc} test. The upper triangle shows difference between intersections. Lower triangle shows pairs with statistical significance.}
\label{tab:nem}
\begin{tabular}{l c c c}
\toprule

\textbf{Compared Measures}	&	\textbf{Degree\ $\cap$ ARL} 	&	\textbf{Page Rank $\cap$ ARL}	&	\textbf{Page Rank $\cap$ Degree}\\

\hline
Degree\ $\cap$ ARL  		&	- 	&	1.00	&	2.00 \\
Page Rank $\cap$ ARL		&	-  	&	-	    &	1.00 \\
Page Rank $\cap$ Degree 		& $^*$, $^{**}$	&	-	&	- \\
\bottomrule
\multicolumn{4}{c}{\footnotesize$^*$ significant at $p<0.05$, CD: $1.253$; $^{**}$ significant at $p<0.01$, CD: $1.557$.} \\
\end{tabular}
\end{table}
The three different methods were investigated to identify influential users.
The amount of time needed to identify influential users differs between the methods. This is shown in Table~\ref{tab:snaTime}. Rule based learning is suggested to be the fastest method, and Page Rank the slowest. This might be explained by Page Rank being a global measure compared to the Degree, which is a local measure.
The execution time of the different methods with the confidence intervals are also presented in Figure~\ref{fig:sna_time}, where intuitively it would seem that the rule based learner has a significantly lower execution time than the other methods.

\begin{table}[H]
\caption{Mean execution time for ranking users.}
\label{tab:snaTime}
\centering
\begin{tabular}{l r r}
\toprule

\textbf{Method} & \textbf{Mean} & \textbf{Std.} \\

\hline
Degree & 329.135 & (2345.996) \\
Page Rank & 633.152 & (4602.607) \\
ASR & 9.033 & (22.497) \\
\bottomrule
\end{tabular}
\end{table}

Whether there is any statistical significant difference is evaluated using a Kruskal--Wallis test followed by a pair-wise Wilcoxon \textit{post hoc} test~\cite{Sheskin:2007tx}.
The Kruskal--Wallis test is used to see if there is a significant difference between any of the methods, and the \textit{post hoc} test is used to detect between which methods the differences exist.
The Kruskal--Wallis test detected a significant difference between the methods ($\chi^{2}=6.626$, $df=2$, \emph{p} < 0.05). The Wilcoxon \textit{post hoc} tests showed a significant difference between Rule based and Degree (\emph{p} < 0.05, $w = 14130$). No other statistical significant differences were found. While there exists a large difference in mean, there is no detectable significant difference between the Association Rule based method and Page Rank($p = 0.054$, $w = 13704$). This might be due to the high standard deviation.

\section{Discussion} 
\label{sec:discussion}

Users within online social networks create a large amount of generated data in the form of interactions (comments and likes). Not enough attention has been put on the analysis of how users influence each other and how to predict the behavior of users within Facebook groups. In this paper, we have collected a significant amount of user data and then by using association rule learning, implemented and examined how users influence each other. Based on the results and analysis, we are able to determine to what extent users influence other users to participate and interact in \mbox{new groups.}

To verify the results from the page Occupy\-Together, an additional 195 pages were sampled to verify our assumptions.
These pages were reduced to 108 due to size constraints. Arguably, pages that were too large could have been processed by limiting the time span, \emph{i.e}., instead of considering all six years of the page, a time span of the latest six months could have been considered. Association rules were computed for each page in our dataset.  For association rules with confidence~$\geqslant$95\%, the mean was $33,426.89\ (sd=87,457.39)$, and a median of $2351$ was found for the number of rules.

The computed rules were tested resulting in an average of $0.913\ (sd=0.115)$ for accuracy, $0.614\ (sd=0.340)$ for precision, and $0.141\ (sd=0.256)$ for recall when predicting user activity on a post. In other words, it is possible to predict a subset of users' future participation with \mbox{high correctness.}

The results also indicate that influential users can be identified using association rule learning. That is, users on the left-hand-side, in a rule with high confidence and high lift, are influencing users on the right-hand-side to participate in the conversation. These results have been verified and compared with the traditional network analysis methods, Page Rank Centrality and Degree Centrality. Showing that at best $\sim$30\% of the users ranked using association rule learning overlap with the users ranked using traditional methods.

Interestingly, association rule learning are magnitudes faster in execution time for ranking users than other methods. Another finding related to the ranking of users is that we see no significant difference between ranked influential users based on Page Rank or Degree. However, we show that Page Rank is a more time consuming algorithm.

The main disadvantage of association rule learning is the fact that we cannot extract rules for the biggest pages in our dataset. We have not shown in this paper that association rule learning is better/or worse than other approaches. However, it was not the point of our research. Since there is no ground truth, it is not possible to say which approach is better (or worse). Our objective was to present a different approach for identifying influential users and leave the final decision of which approach to use to the researcher.

Furthermore, from the list of influential users, presented in Section~\ref{sec:sna}, it is also possible to limit the size of the item-set. This will result in an increasing speed when building rules without a significant decrease in quality of the rules.
As a validation threat, information on Facebook is filtered by a secret algorithm. This poses a potential validity threat to our results as users are presented posts filtered by the algorithm. For example, a reason for a user not commenting on a post might be due to visibility (the filtering algorithm is not presenting the post to the user) rather than by topic.

\section{Conclusions}
\label{sec:conclusion}

This article presents four contributions. Firstly, insights on user behavior on public pages on Facebook indicates that the top 10\% and top 20\% of users corresponds to a vast majority of the content. Secondly, it is possible to identify influential users using association rule learning. The results indicate no statistically significant difference between our rule based method compared to Page Rank. Thirdly, execution times of well known methods for ranking users in social media together with our approach using association rule learning are investigated. The results suggest that rule based ranking of users has lower execution time compared to state-of-the-art methods, $9.0$ \emph{vs}. $633.1$ and $329.1$ seconds on average.
Finally, the article verifies how association rule learning can be used to predict user participation in social media pages on Facebook. The results indicate an average prediction accuracy of $0.913\ (sd=0.115)$ for the association rule learning approach.

For future work, it would be interesting to investigate rule creation with a time series perspective of the data e.g., using a sliding window approach. Additionally, methods to investigate a subset of users for rule creation need to be investigated.

\vspace{6pt}  


\acknowledgments{\textbf{Acknowledgments:} This work was partially supported by the European Union's Seventh Framework Program for research, technological development and demonstration under grant agreement No. 316097 [ENGINE] and by The Polish National Science Center, decision No. DEC-2013/09/B/ST6/02317}


\authorcontributions{\textbf{Author Contributions:} Fredrik Erlandsson and Piotr Br\'odka conceived and designed the experiments; \mbox{Fredrik Erlandsson} performed the experiments; Fredrik Erlandsson, Piotr Br\'odka and Anton Borg analyzed the data; Henric Johnson enabled the work and also contributed with critical revision. All authors have written, read, and approved the final manuscript.}


\conflictofinterests{\textbf{Conflicts of Interest:} The authors declare no conflict of interest.}


\newpage
\bibliographystyle{mdpi}

\begin{thebibliography}{999}

\bibitem[Cha \em{et~al.}(2010)Cha, Haddadi, Benevenuto, and
  Gummadi]{cha2010measuring}
Cha, M.; Haddadi, H.; Benevenuto, F.; Gummadi, P.K.
\newblock Measuring User Influence in Twitter: The Million Follower Fallacy.
\newblock {\em ICWSM} {\bf 2010}, {\em 10}, 10--17.

\bibitem[Riquelme(2015)]{DBLP:journals/corr/Riquelme15}
Riquelme, F.
\newblock Measuring user influence on Twitter: {A} survey.
\newblock {\bf 2015}, {arXiv:1508.07951}.

\bibitem[Musia\l \em{et~al.}(2009)Musia\l, Kazienko, and
  Br\'{o}dka]{Musial:2009:UPM:1731011.1731017}
Musia\l, K.; Kazienko, P.; Br\'{o}dka, P.
\newblock User Position Measures in Social Networks.
\newblock  In \emph{Proceedings of the 3rd Workshop on Social Network Mining and
  Analysis}; ACM: New York, NY, USA, 2009; Article No. 6.

\bibitem[Br{\'o}dka(2013)]{brodka2013key}
Br{\'o}dka, P.
\newblock Key User Extraction Based on Telecommunication Data (aka. Key Users
  in Social Network. How to find them?).
\newblock  {\bf 2013}, {arXiv:1302.1369}.

\bibitem[Erlandsson \em{et~al.}(2016)Erlandsson, Borg, Johnson, and
  Br{\'o}dka]{Erlandsson:2016kq}
Erlandsson, F.; Borg, A.; Johnson, H.; Br{\'o}dka, P.
\newblock {Predicting User Participation in Social Media}. In {\em Advances in
  Network Science}; Springer International Publishing: Cham, Switserland, 2016; pp.
  126--135.

\bibitem[Flach(2012)]{flach2012machine}
Flach, P.
\newblock {\em {Machine Learning: The Art and Science of Algorithms that Make
  Sense of Data}}; Cambridge University Press:  Cambridge, UK, 2012.

\bibitem[Liben-Nowell and
  Kleinberg(2007)]{Liben-Nowell:2007:LPS:1241540.1241551}
Liben-Nowell, D.; Kleinberg, J.
\newblock The Link-prediction Problem for Social Networks.
\newblock {\em J. Am. Soc. Inf. Sci. Technol.} {\bf 2007}, {\em
  58},~1019--1031.

\bibitem[Utz and Jankowski(2015)]{utz2015making}
Utz, S.; Jankowski, J.
\newblock Making ``Friends'' in a Virtual World The Role of Preferential
  Attachment, Homophily, and Status.
\newblock {\em Soc. Sci. Comput. Rev.} {\bf 2015}, doi:10.1177/0894439315605476S.

\bibitem[Zu \em{et~al.}(2015)Zu, Hu, Gu, and Seng]{xia}
Zu, Q.; Hu, B.; Gu, N.; Seng, S.
\newblock Human Centered Computing. In Proceedings of the 1st Human Centered Computing Conference International Conference, {(HCC 2014)},
  Phnom Penh, Cambodia, 27--29 November 2014.

\bibitem[Au \em{et~al.}(2003)Au, Chan, and Yao]{au2003novel}
Au, W.H.; Chan, K.C.; Yao, X.
\newblock A novel evolutionary data mining algorithm with applications to churn
  prediction.
\newblock {\em IEEE Trans. Evolut. Comput.} {\bf 2003}, {\em
  7},~532--545.

\bibitem[Ruta \em{et~al.}(2009)Ruta, Kazienko, and Br{\'o}dka]{ruta2009network}
Ruta, D.; Kazienko, P.; Br{\'o}dka, P.
\newblock Network-Aware Customer Value in Telecommunication Social Networks.
\newblock  In Proceedings of the 2009 International Conference on Artificial Intelligence, (ICAI'09), Las Vegas, NE, USA,  13--16 July 2009; pp. 261--267.

\bibitem[Saganowski \em{et~al.}(2015)Saganowski, Gliwa, Bródka, Zygmunt,
  Kazienko, and Koźlak]{saganowski2015predicting}
Saganowski, S.; Gliwa, B.; Br\'{o}dka, P.; Zygmunt, A.; Kazienko, P.; Ko\'{z}lak, J.
\newblock Predicting community evolution in social networks.
\newblock {\em Entropy} {\bf 2015}, {\em 17},~3053--3096.

\bibitem[De~Meo \em{et~al.}(2015)De~Meo, Ferrara, Rosaci, and
  Sarne]{DeMeo:2015cf}
De~Meo, P.; Ferrara, E.; Rosaci, D.; Sarne, G.M.L.
\newblock {Trust and Compactness in Social Network Groups}.
\newblock {\em IEEE Trans. Cybern.} {\bf 2015}, {\em
  45},~205--216.

\bibitem[Asur and Huberman(2010)]{Asur:2010:PFS:1913793.1914092}
Asur, S.; Huberman, B.A.
\newblock Predicting the Future with Social Media.
\newblock  In \emph{Proceedings of the 2010 IEEE/WIC/ACM International Conference on Web
  Intelligence and Intelligent Agent Technology--Volume 01}; IEEE Computer
  Society: Washington, DC, USA,  2010; pp. 492--499.

\bibitem[Ahmad \em{et~al.}(2010)Ahmad, Riaz, Johnson, and Lavesson]{Ahmad2010}
Ahmad, W.; Riaz, A.; Johnson, H.; Lavesson, N.
\newblock Predicting Friendship Intensity in Online Social Networks.
\newblock  In \emph{Proceedings of the 21st Tyrrhenian Workshop on Digital
  Communications: Trustworthy Internet}; Springer: Berlin/Heidelberg, Germany, 2010.

\bibitem[Nia \em{et~al.}(2013)Nia, Erlandsson, Johnson, and
  Wu]{nia2013leveraging}
Nia, R.; Erlandsson, F.; Johnson, H.; Wu, S.F.
\newblock Leveraging social interactions to suggest friends.
\newblock   In Proceedings of the 2013 IEEE 33rd
  International Conference on Distributed Computing Systems Workshops (ICDCSW), Philadelphia, PA, USA, 8--11 July 2013; pp. 386--391.

\bibitem[Spertus \em{et~al.}(2005)Spertus, Sahami, and
  Buyukkokten]{Spertus:2005ec}
Spertus, E.; Sahami, M.; Buyukkokten, O.
\newblock { {Evaluating Similarity Measures: A Large-Scale Study in the
  Orkut Social Network}}. In Proceedings of the Eleventh ACM SIGKDD International Conference on Knowledge Discovery in Data Mining, (KDD'05), Chicago, IL, USA, 21--24 August 2005; pp.~678--684.

\bibitem[Vasuki \em{et~al.}(2011)Vasuki, Natarajan, Lu, Savas, and
  Dhillon]{Vasuki:2011ku}
Vasuki, V.; Natarajan, N.; Lu, Z.; Savas, B.; Dhillon, I.
\newblock {Scalable Affiliation Recommendation Using Auxiliary Networks}.
\newblock {\em ACM Trans. Intell. Syst. Technol.}
  {\bf 2011}, {\em 3}, doi:10.1145/2036264.2036267.

\bibitem[Petz \em{et~al.}(2015)Petz, Karpowicz, F{\"u}rschu{\ss}, and
  Auinger]{Petz:2015fj}
Petz, G.; Karpowicz, M.; F{\"u}rschu{\ss}, H.; Auinger, A.
\newblock {Reprint of: Computational approaches for mining user's opinions on
  the Web 2.0}.
\newblock {\em Inf. Process.} {\bf 2015}, {\em 51},~510--519.

\bibitem[Jamali and Rangwala(2009)]{5368318}
Jamali, S.; Rangwala, H.
\newblock Digging Digg: Comment Mining, Popularity Prediction and Social
  Network Analysis.
\newblock  In Proceedings of the International
  Conference on Web Information Systems and Mining, (WISM~2009), {Shanghai, China, 7--8 November} 2009; pp. 32--38.

\bibitem[Hakim and Khodra(2014)]{7062665}
Hakim, M.; Khodra, M.
\newblock Predicting information cascade on Twitter using support vector
  regression.
\newblock  In {Proceedings of the 2014 International
  Conference on Data and Software Engineering (ICODSE}),  {Hyderabad, India, 31 May--7 June} 2014; pp. 1--6.

\bibitem[Jankowski \em{et~al.}(2012)Jankowski, Michalski, and
  Kazienko]{Jankowski2012}
Jankowski, J.; Michalski, R.; Kazienko, P.
\newblock The Multidimensional Study of Viral Campaigns as Branching Processes.
  In {\em Social Informatics}; Aberer, K., Flache, A., Jager, W., Liu, L.,
  Tang, J., Gu\'{e}ret, C., Eds.; Springer: Berlin/Heidelberg,  Germany, 2012; Volume 7710,
  pp. 462--474.

\bibitem[Bakshy \em{et~al.}(2011)Bakshy, Hofman, Mason, and
  Watts]{Bakshy:2011go}
Bakshy, E.; Hofman, J.M.; Mason, W.A.; Watts, D.J.
\newblock {{Everyone's an Influencer: Quantifying Influence on Twitter.}} In {\em Proceedings of the Fourth ACM International Conference on Web Search and Data Mining, (WSDM~'11)};
ACM: New York, NY, USA, 2011; pp.~65--74.

\bibitem[Ghosh and Lerman(2010)]{Ghosh:2010vz}
Ghosh, R.; Lerman, K.
\newblock {Predicting Influential Users in Online Social Networks}.
\newblock {\bf 2010}, arXiv:1005.4882.


\bibitem[Shin \em{et~al.}(2008)Shin, Xu, and Kim]{Shin:2008cz}
Shin, H.; Xu, Z.; Kim, E.Y.
\newblock {Discovering and Browsing of Power Users by Social Relationship Analysis in Large-Scale Online Communities}. In {\em Proceedings of the 2008 IEEE/WIC/ACM International Conference on Web Intelligence and Intelligent Agent Technology--Volume 01};  {IEEE Computer Society: Washington, DC, USA}, 2008; pp.~105--111.

\bibitem[Lin \em{et~al.}(2015)Lin, Wu, Chen, and Yang]{Lin:2015kh}
Lin, K.C.; Wu, S.H.; Chen, L.P.; Yang, P.C.
\newblock {Finding the Key Users in Facebook Fan Pages via a Clustering
  Approach}.
\newblock {In Proceedings of the 2015 IEEE International
  Conference on Information Reuse and Integration (IRI)}, {Redwood City, CA, USA, 13--15 August 2015}; pp. 556--561.

\bibitem[Weng \em{et~al.}(2010)Weng, Lim, Jiang, and He]{Weng:2010fd}
Weng, J.; Lim, E.P.; Jiang, J.; He, Q.
\newblock {TwitterRank: Finding Topic-Sensitive Influential Twitterers}. In {\em Proceedings of the Third ACM International Conference on Web Search and Data Mining}; ACM:  New York, NY, USA, 2010; pp.~261--270.

\bibitem[Tang and Yang(2010)]{Tang:2010dh}
Tang, X.; Yang, C.C.
\newblock {Identifing Influential Users in an Online Healthcare Social
  Network}. In Proceedings of 2010 IEEE International Conference on Intelligence and Security Informatics (ISI), Vancouver, BC, Canada ,23--26 May 2010; pp.~43--48.

\bibitem[Hotho \em{et~al.}(2006)Hotho, J{\"a}schke, Schmitz, and
  Stumme]{Hotho:2006bs}
Hotho, A.; J{\"a}schke, R.; Schmitz, C.; Stumme, G.
\newblock {Information Retrieval in Folksonomies: Search and Ranking}. In {\em
  The Semantic Web: Research and Applications}; Springer: Berlin/Heidelberg, Germany,
  2006; pp. 411--426.

\bibitem[Nancy \em{et~al.}(2013)Nancy, Geetha~Ramani, and Jacob]{Nancy2013}
Nancy, P.; Geetha~Ramani, R.; Jacob, S.
\newblock Mining of Association Patterns in Social Network Data (Face Book 100
  Universities) through Data Mining Techniques and Methods. In {\em Advances in
  Computing and Information Technology}; Meghanathan, N., Nagamalai, D., Chaki,
  N., Eds.; Springer: Berlin/Heidelberg,  Germany, 2013; Volume 178,  pp. 107--117.

\bibitem[Yu \em{et~al.}(2014)Yu, Liu, Shi, Hwang, Wan, and Lu]{Xiaoqing}
Yu, X.; Liu, H.; Shi, J.; Hwang, J.N.; Wan, W.; Lu, J.
\newblock Association Rule Mining of Personal Hobbies in Social Networks.
\newblock  In Proceedings of the 2014 IEEE International Congress on Big Data (BigData Congress),  { Anchorage, AK, USA, 27 June 27--2 July}  2014; pp. 310--314.

\bibitem[Schmitz \em{et~al.}(2006)Schmitz, Hotho, J{\"a}schke, and
  Stumme]{Schmitz2006}
Schmitz, C.; Hotho, A.; J{\"a}schke, R.; Stumme, G.
\newblock Mining association rules in folksonomies. In {\em Data Science and
  Classification}; Springer: Berlin/Heidelberg, Germany,  2006; pp. 261--270.

\bibitem[Agrawal \em{et~al.}(1993)Agrawal, Imieli\'{n}ski, and
  Swami]{Agrawal:1993:MAR:170036.170072}
Agrawal, R.; Imieli\'{n}ski, T.; Swami, A.
\newblock Mining Association Rules Between Sets of Items in Large Databases.
\newblock {\em ACM SIGMOD Rec.} {\bf 1993}, {\em 22},~207--216.

\bibitem[Agrawal and Srikant(1994)]{Agrawal:1994:FAM:645920.672836}
Agrawal, R.; Srikant, R.
\newblock Fast Algorithms for Mining Association Rules in Large Databases.
\newblock  In \emph{Proceedings of the 20th International Conference on Very Large Data
  Bases}; Morgan Kaufmann Publishers Inc.: San Francisco, CA, USA,  1994; pp.
  487--499.

\bibitem[Goethals(2003)]{Goethals02surveyon}
Goethals, B.
\newblock {\em Survey on Frequent Pattern Mining};
\newblock Technical report; University of Helsinki: Helsinki, Finland, 2003.

\bibitem[Zaki(2000)]{10.1109/69.846291}
Zaki, M.J.
\newblock Scalable Algorithms for Association Mining.
\newblock {\em IEEE Trans. Knowl. Data Eng.} {\bf 2000},
  {\em 12},~372--390.

\bibitem[Erlandsson \em{et~al.}(2015)Erlandsson, Nia, Boldt, Johnson, and
  Wu]{erlandsson2015crawling}
Erlandsson, F.; Nia, R.; Boldt, M.; Johnson, H.; Wu, S.F.
\newblock Crawling Online Social Networks.
\newblock  In Proceedings of the 2015 European Network Intelligence Conference (ENIC), {Karlskrona, Sweden, 21--22 September }  2015.

\bibitem[Nia \em{et~al.}(2012)Nia, Erlandsson, Bhattacharyya, Rahman, Johnson,
  and Wu]{nia2012sin}
Nia, R.; Erlandsson, F.; Bhattacharyya, P.; Rahman, M.R.; Johnson, H.; Wu, S.F.
\newblock Sin: A platform to make interactions in social networks accessible.
\newblock  In Proceedings of the 2012 International
  Conference on Social Informatics (SocialInformatics),  {Washington, DC, USA, 14--16 December } 2012; pp. 205--214.

\bibitem[]{OccupyTogether}
\newblock{Occupy Together. Available online: https://www.facebook.com/OccupyTogether} (accessed on 27 April 2016).

\bibitem[Brodka \em{et~al.}(2009)Brodka, Musial, and
  Kazienko]{brodka2009performance}
Brodka, P.; Musial, K.; Kazienko, P.
\newblock A performance of centrality calculation in social networks.
\newblock   In Proceedings of the International Conference on IEEE Computational Aspects of Social Networks (CASON'09), Fontainebleau, France, 24--27 June 2009; pp. 24--31.

\bibitem[Sheskin(2007)]{Sheskin:2007tx}
Sheskin, D.
\newblock {\em {Handbook of Parametric and Nonparametric Statistical
  Procedures}}; Chapman {\&} Hall:  London, UK,~2007.

\bibitem[Dem{\v s}ar(2006)]{Demsar:2006un}
Dem{\v s}ar, J.
\newblock {Statistical Comparisons of Classifiers over Multiple Data Sets}.
\newblock {\em  J. Mach. Learn. Res.} {\bf 2006}, {\em
  7},~1--30.



\end{thebibliography}
\renewcommand\bibname{References}


%


%

\end{document}